\documentclass[aps,prd,twocolumn,preprintnumbers,nofootinbib]{revtex4-1}
\usepackage{amsmath}
\usepackage{amsfonts}
\usepackage{graphicx}
\usepackage[config=altsf]{subfig}
\usepackage{feynmp}
\usepackage{color}
\usepackage{dsfont}
\usepackage{array}
\usepackage{slashed}
\usepackage{afterpage}
\usepackage{appendix}
\usepackage{multirow}
\def\lesssim{\mathrel{\mathpalette\vereq<}}

\begin{document}

\title{Skew-Flavored Dark Matter}

\author{Prateek Agrawal}
\affiliation{Fermilab, P.O. Box 500, Batavia, IL 60510, USA}
\affiliation{Department of Physics, Harvard University, Cambridge, MA,
02138, USA}
\author{Zackaria Chacko}
\affiliation{Maryland Center for Fundamental Physics, Department of Physics, 
University of Maryland, College Park, MD 20742, USA}
\author{Elaine C. F. S. Fortes}
\affiliation{1 Instituto de Fisica Teorica, Universidade Estadual Paulista, 
Rua Dr. Bento Teobaldo Ferraz 271, 01140-070 Sao Paulo, SP, Brazil}
\author{Can Kilic}
\affiliation{Theory Group, Department of Physics and Texas Cosmology Center,
The University of Texas at Austin, Austin, TX 78712, USA}

\begin{abstract}

We explore a novel flavor structure in the interactions of dark matter 
with the Standard Model. We consider theories in which both the dark 
matter candidate, and the particles that mediate its interactions with 
the Standard Model fields, carry flavor quantum numbers. The 
interactions are skewed in flavor space, so that a dark matter particle 
does not directly couple to the Standard Model matter fields of the same 
flavor, but only to the other two flavors. This framework respects 
Minimal Flavor Violation, and is therefore naturally consistent with 
flavor constraints. We study the phenomenology of a benchmark model in 
which dark matter couples to right-handed charged leptons. In large 
regions of parameter space the dark matter can emerge as a thermal 
relic, while remaining consistent with the constraints from direct and 
indirect detection. The collider signatures of this scenario include 
events with multiple leptons and missing energy. These events exhibit a 
characteristic flavor pattern that may allow this class of models to 
be distinguished from other theories of dark matter.

\end{abstract}

\preprint{UTTG-20-15 , TCC-009-15}

\maketitle

\section{Introduction}

It is now well established that about four-fifths of the matter density 
of the universe is made up of some form of dark matter (DM) that is 
neutral under both the color and electromagnetic interactions, and that 
lies outside the Standard Model (SM) of particle physics. However, the 
precise nature of the particles of which DM is composed continues to 
remain a mystery. One well-motivated possibility is that DM is composed 
of ``Weakly Interacting Massive Particles" (WIMPs), particles with 
masses of order the weak scale that have interactions of weak scale 
strength with the Standard Model (SM) fields. The WIMP scenario is 
rather appealing because, provided the WIMPs were in equilibrium with 
the SM at early times, just enough of them survive today as thermal 
relics to account for the observed density of DM.

An attractive feature of the WIMP framework is that the weak scale 
strength interaction of DM with the SM fields can be probed at current 
experiments.  Nevertheless, efforts to produce DM at high energy 
colliders have thus far proven
unsuccessful~\cite{Khachatryan:2014rra,Aad:2015zva}. Likewise,
direct detection 
experiments, which seek to observe DM in the laboratory through its 
scattering off nucleons, have also been fruitless~\cite{Akerib:2015rjg}.
Indirect detection 
experiments do offer some tantalizing hints of signals that may 
originate from the annihilation of
DM~\cite{Goodenough:2009gk,Daylan:2014rsa,Boyarsky:2014jta,Bulbul:2014sua},
but these have not been confirmed 
so far.

New physics at the weak scale is severely constrained by flavor 
experiments. Since any WIMP DM candidate is required to have 
interactions of order weak scale strength with the SM fields, any flavor 
violating couplings of DM with the SM fields are constrained to be 
extremely small. This constitutes a severe restriction on this class of 
theories. We shall refer to this as the ``WIMP Flavor Problem". In the 
class of theories in which the relic abundance of DM is set by 
annihilation into the weak gauge bosons of the SM as, for example, in 
Minimal Dark Matter~\cite{Cirelli:2005uq}, or through the Higgs 
portal~\cite{Silveira:1985rk,McDonald:1993ex,Burgess:2000yq}, this 
problem can naturally be avoided. However, in theories in which the 
relic abundance is set by direct couplings of the DM candidate to the SM 
matter fields, these interactions are required to be closely aligned 
with the SM Yukawa couplings.

In supersymmetric theories with neutralino or sneutrino DM, the WIMP 
Flavor Problem is generally subsumed into the well-known 
``Supersymmetric Flavor Problem"~\cite{Gabbiani:1996hi}. Solutions to 
the Supersymmetric Flavor Problem, such as gauge 
mediation~\cite{Dine:1981za,Dimopoulos:1981au, 
AlvarezGaume:1981wy,Dine:1993yw,Dine:1994vc,Dine:1995ag,Giudice:1998bp}, 
anomaly mediation~\cite{Randall:1998uk,Giudice:1998xp} and gaugino 
mediation~\cite{Kaplan:1999ac,Chacko:1999mi} then automatically resolve 
the WIMP Flavor Problem. The WIMP Flavor Problem may be thought of as 
the generalization of the Supersymmetric Flavor Problem to the larger 
class of theories of WIMP DM.

The phenomenology of a WIMP DM model depends sensitively on the flavor 
structure associated with the couplings of the DM candidate to the SM, 
which in turn on how the WIMP Flavor Problem is solved.  It is therefore 
very important to identify flavor structures that naturally resolve the 
WIMP Flavor Problem. One convenient framework that is sufficient, 
although not necessary, to suppress flavor violation is Minimal Flavor 
Violation (MFV)~\cite{Chivukula:1987py, Hall:1990ac,Buras:2000dm, 
D'Ambrosio:2002ex,Buras:2003jf}. Within this framework, the flavor 
structure of the entire theory is set by the SM Yukawa couplings. This 
automatically ensures that the flavor violation pattern conforms to that 
of the SM, making it compatible with experiment.  A simple way to 
implement MFV is to elevate the SM Yukawa couplings to spurions that 
transform under the flavor group and require the theory to be invariant 
under these spurious flavor symmetries. Most commonly, the DM has been 
assumed to be a singlet under the SM flavor symmetries. More recently, 
realizations of MFV in which the DM particle carries flavor quantum 
numbers and transforms under the flavor group have been 
explored~\cite{Batell:2011tc,Agrawal:2011ze,Kile:2013ola}.

At present, the status of experimental observations provides additional
motivation for considering theories in which DM comes in multiple 
flavors~\cite{MarchRussell:2009aq,Batell:2011tc, Agrawal:2011ze, Kumar:2013hfa, 
Lopez-Honorez:2013wla, Kile:2013ola, Batell:2013zwa, Agrawal:2014una, 
Agrawal:2014aoa, Hamze:2014wca, Lee:2014rba,Kilic:2015vka, 
Agrawal:2015tfa,
Bishara:2015mha,Bhattacharya:2015xha,
Calibbi:2015sfa,Baek:2015fma,Chen:2015jkt}, or has
flavor-specific couplings to 
the SM~\cite{Kile:2011mn, Kamenik:2011nb, Zhang:2012da, Kile:2014jea}. 
The absence of signals in direct detection implies that the couplings of 
DM to the first generation SM quarks must be small. Theories in which the 
DM particle couples preferentially to the heavier quark
flavors~\cite{Cheung:2010zf,Agrawal:2011ze,Kumar:2013hfa,Zhang:2012da} can 
naturally satisfy this bound, while remaining consistent with the relic 
abundance constraint. In addition, flavored DM has been proposed as an 
explanation of the gamma ray excess observed emanating from the galactic 
center \cite{Agrawal:2014una,Agrawal:2014aoa}, and also the observed 3.5 
keV line in the X-ray spectrum~\cite{Agrawal:2015tfa}.

In this paper, we explore a novel flavor structure for the coupling of 
DM to the SM that is consistent with MFV. We focus on theories in which 
the DM has renormalizable contact interactions with SM matter and a 
mediator. If all three fields transform as the fundamental 
representation of the SU(3) flavor group associated with the SM field, 
it is possible to construct flavor invariants using the completely 
antisymmetric $\epsilon_{ijk}$ contraction, leading to interactions with 
a ``skewed texture''.

As an example, consider the case in which the DM particle, a SM
singlet Dirac fermion which we label as $\chi$, and the mediator, a
complex scalar which we label as $\phi$ interact with the right-handed
SM leptons, with all three fields transforming in the fundamental
representation of ${\rm SU(3)}_{E}$, the flavor symmetry of the
right-handed leptons. The interaction term has the form
 \begin{align}
\lambda^{ijk} \chi_i E^c_j \phi_k \; \; + {\rm h.c.}
 \end{align}
 In this expression $i$, $j$ and $k$ represent SU(3$)_{E}$ quantum 
numbers and define the flavor labels of the DM and the mediator with
respect to charged leptons. Then, if the theory is to be consistent
with MFV, the flavor structure associated with this interaction must
respect the SU(3$)_{E}$ flavor symmetry, up to effects associated with
the SM Yukawa couplings.

If we write the lepton Yukawa couplings of the SM as
\begin{align}
  {y_A}^i L^A E^c_i H \; \; + {\rm h.c.},
\end{align}
the Yukawa matrix ${y_A}^i$ can be thought of as a spurion
transforming as $(3,\bar{3})$ under the SU(3$)_L \times $ SU(3$)_E$
flavor symmetry.
Then, if the theory respects
MFV the matrix $\lambda$ is restricted to be of the form 
\begin{equation}
  {\lambda}^{ijk} = \alpha \epsilon^{ijk} + \delta {\lambda}^{ijk}
\end{equation}
where $\delta {\lambda}^{ijk}$ is given by
\begin{equation}
  \beta_1 \; {\left(y^{\dagger} y \right)_{i'}}^{i} \epsilon^{i'jk}
  +  \beta_2 \; {\left(y^{\dagger} y \right)_{j'}}^{j} \epsilon^{ij'k}
  + \beta_3 \; {\left(y^{\dagger} y \right)_{k'}}^{k} \epsilon^{ijk'} \; .
  \label{genint}
\end{equation}
Here the $\alpha$ and $\beta$ parameters are constants, and we are 
keeping only the leading terms in an expansion in powers of the SM 
Yukawa couplings. We see from this that the couplings of the DM particle 
are skewed in flavor space, so that, for example, $\chi_{\tau}$ couples 
to the electron and the muon, but not to the tau. Nevertheless, because 
this construction respects MFV, it is consistent with the bounds on 
flavor violating processes.

We can write the DM mass term schematically as
\begin{equation}
  \left[ m_{\chi} \right]{_i}^j \bar{\chi}^{i} \chi_{j} \;.
\end{equation}
In this case MFV restricts $m_{\chi}$ to be of the form
\begin{align}
  \left[ m_{\chi} \right]{_i}^j
  = {\left( m_0 \mathds{1} + \Delta m \; y^{\dagger} y \right)_i}^j \; , 
\end{align} 
 where $m_0$ and $\Delta m$ are constants. Since the SM Yukawa couplings 
are small, the various DM flavors are expected to have small splittings. 
Either the tau flavored or the electron flavored state will be the 
lightest, depending on the sign of $\Delta m$.

Similarly, we can write the mediator mass term schematically as
\begin{equation}            
  \left[ \hat{m}^{2}_{\phi} \right]{_i}^j {\phi^\dagger}^{i} \phi_{j} \;.
\end{equation}
As before, MFV restricts $\hat{m}^{2}_{\phi}$ to be of the form
\begin{align}
  \left[ \hat{m}^{2}_{\phi} \right]{_i}^j
  = {\left( m^{2}_\phi \mathds{1} + \Delta m^{2}_\phi \; y^{\dagger} y \right)_i}^j \; ,
\end{align}
where $m^{2}_\phi$ and $\Delta m^{2}_\phi$ are constants. Once again,
we expect that the splittings will be small. 

In the next section we consider in detail a simple benchmark model that 
realizes the skew-flavored DM scenario, in which DM couples to the 
right-handed leptons of the SM. We determine the range of parameters 
that leads to the observed abundance of DM, and study the implications 
for direct detection, indirect detection and for the LHC. We then 
briefly consider alternative realizations of skew-flavored DM, in which 
DM couples to the left-handed leptons or quarks, before concluding.

\section{A Benchmark Model}

In this section we consider in detail a benchmark DM model that exhibits 
skewed flavor structure.\footnote{In accordance with the WIMP nature of our model, we will take the mass scale of the DM particles $\chi$ and mediators $\phi$ to be in the GeV-TeV range, with $m_{\chi}<m_{\phi}$.} We restrict the interaction term to be of the form 
 \begin{equation}
\lambda\; \epsilon^{ijk} \chi_i E^c_j \phi_k \; \; + {\rm h.c.}
\label{epsint}
 \end{equation}
 This corresponds to the limit in which the flavor non-universal terms 
in Eq.~(\ref{genint}) are small, and can be neglected. Expanding this out
in the language of 4-component Dirac fermions, the couplings 
of the electron flavor of DM take the form,
\begin{equation}
  \mathcal{L}
  \supset
  \lambda
  \left[
    \bar{\chi}_e
    \frac{1+\gamma_5}{2}\left(\mu \;\phi_{\tau} - \tau \;\phi_{\mu} \right) 
  \right] 
  +{\rm h.c.}
  \label{lepton4comp}
\end{equation}
The couplings of the $\mu$ and $\tau$ flavors of DM may be obtained by 
cyclically permuting the $e$, $\mu$ and $\tau$ flavors in the 
expression above. This is to be contrasted with the form of the coupling 
in the more conventional realization of flavored DM, where the mediator 
$\phi$ is a flavor singlet,
 \begin{equation}
\lambda\; \chi^i E^c_i \phi \; \; + {\rm h.c.}
\label{convFDM}
 \end{equation}

 As regards the masses of the DM particles, we work in the limit that 
the $e$ and $\mu$ flavors of DM are very nearly degenerate, so that 
their splitting can be neglected. However, because of the much larger 
$\tau$ Yukawa coupling, we allow for the possibility that the $\tau$ 
flavor of DM is somewhat split from the other two. Accordingly, we write 
the DM mass term as
\begin{equation}
  m_0 \left( \bar{\chi}_e \chi_e + \bar{\chi}_{\mu} \chi_{\mu} +
  \bar{\chi}_{\tau} \chi_{\tau} \right) +
  \Delta m \bar{\chi}_{\tau} \chi_{\tau}
  \; .
\end{equation}
The sign of $\Delta m$ determines which flavor of DM is the lightest.  
In the limit that $\Delta m$ is very small, all three DM flavors are 
nearly degenerate. It is important to note, however, that even fairly 
small splittings $\Delta m$ cannot always be neglected, since the 
lifetimes of the heavier flavors depend on this splitting.

Small splittings in the mediator masses, however, do not significantly 
affect the phenomenology. Therefore, for simplicity, we take all the 
different flavors of mediator to have a common mass $m_\phi$. In the 
rest of this paper, we will focus our attention on two mass hierarchies, 
one where $\chi_{\tau}$ is significantly lighter than $\chi_{e,\mu}$ 
such that it freezes out after the other two have already decayed away, 
and one where all mass splittings are below an MeV so that all three 
$\chi$ flavors are long-lived on cosmological time scales and the relic 
abundance of DM today consists of all three flavors.

\subsection{Relic Abundance}

If DM is a thermal WIMP, its relic abundance is set by its annihilation 
rate to SM states. The primary annihilation mode is through $t$-channel 
$\phi$ exchange to two leptons. We wish to determine the region of 
parameter space that is consistent with the observed abundance of DM.

We first consider the case in which $\chi_\tau$ is the lightest DM 
flavor, and constitutes all of DM. We assume that the mass splitting 
between $\chi_\tau$ and the other flavors is large enough that these 
states decay at early times and play no role in the relic abundance 
calculation. 
 
Since the DM particle is non-relativistic at freeze-out, the 
annihilation processes $\chi_\tau \chi_\tau \rightarrow e^{+} e^{-}$, 
$\mu^+ \mu^-$ proceed through the lowest partial wave. For each of
these final states, only one flavor of $\phi$ is relevant, and
therefore the cross sections add up, leading to
\begin{equation}
  \langle\sigma v\rangle_{relic}
  =
  \frac{2 \lambda^{4} m_\chi^2}{32\pi (m_\chi^2+m_\phi^2)^2}\equiv 2\langle\sigma v\rangle_{0}. \label{eq:annxsecn}
\end{equation} 
Here we have neglected the masses of the final state leptons, and 
defined $\langle \sigma v\rangle_{0}$ as the annihilation cross section 
to just the $e^+ e^-$ (or $\mu^+ \mu^-$) final state. 
Note that the annihilation cross section for the skew-flavored case is 
twice that of the conventional flavored DM case, Eq.~(\ref{convFDM}), 
since in the latter case the only 
annihilation channel is $\chi_\tau \chi_\tau \rightarrow \tau^{+} 
\tau^{-}$.

We next consider the case where all three $\chi$ are degenerate and 
long-lived on cosmological time sales. Taking coannihilations into 
account, we find that the Boltzmann equation for the number density of 
$\chi_e$ is given by, \begin{align}
  \frac{dN_{\chi,e}}{dt}+3H\,N_{\chi,e}
  &= 
  -2\langle \sigma v\rangle_{0} 
  (N^2_{\chi,e}-n_{eq}^2)
 \nonumber \\&\qquad
  -\langle \sigma  v\rangle_{0} 
  (N_{\chi,e}N_{\chi,\mu}-n_{eq}^2)
 \nonumber \\&\qquad
  -\langle \sigma  v\rangle_{0} 
  (N_{\chi,e}N_{\chi,\tau}-n_{eq}^2)
 \nonumber \\&\qquad
 -[\chi_e \to \chi_{\mu,\tau}] \; ,
\end{align} 
 where $n_{eq}$ represents the equilibrium number density for any one 
flavor of DM. The last line includes interactions such as $\chi_e e \to 
\chi_\mu \mu$. These interactions are rapid since the lepton number 
density is not Boltzmann suppressed, and ensure that the number density 
of each flavor of DM is equal. We note that the annihilation cross 
section for two DM particles of the same flavor (such as 
$\bar{\chi}_{e}\chi_{e}$) is twice as large as for a different flavor 
channel (such as either one of $\bar{\chi}_{e}\chi_{\mu}$ or 
$\bar{\chi}_{\mu}\chi_{e}$). Similar equations apply for the number 
densities of $\chi_\mu$ and $\chi_\tau$.

We can sum the Boltzmann equations for each of the three flavors of
$\chi$ to track the total number density of DM, ($N_\chi
\equiv N_{\chi,e} +
N_{\chi,\mu} + N_{\chi_\tau}$),
\begin{align} 
  \frac{dN_{\chi}}{dt}+3H\,N_{\chi}
  &= 
  - \frac{4}{3}\langle \sigma
  v\rangle_{0} \left( N_{\chi}^{2} - \left(3 n_{eq}\right)^{2} \right)
  \, .
\end{align} 
This equation makes it clear that the effective annihilation cross 
section in this case is 
\begin{align}
  \langle \sigma v \rangle_{relic}
  &=
  \frac{4}{3}\langle \sigma v \rangle_{0} \; .
  \label{eq:effxsec3}
\end{align}
This result will also be relevant when we discuss indirect detection.

In Fig.~\ref{fig:exclusion}, we show the parameter space in the 
$m_{\chi}$ vs. $\lambda$ parameter space where the correct relic 
abundance is obtained. We present results for $\phi$ masses of 
$\{200,350,500\}$~GeV and the two mass hierarchies considered above.

\subsection{Direct Detection}

\begin{figure}[tp]
  \begin{center}
    \includegraphics[width=0.3\textwidth]{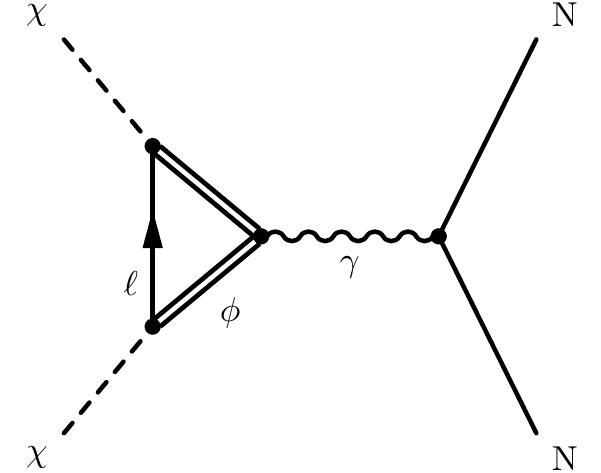}
  \end{center}
  \caption{Feynman diagram contributing to direct detection of dark
  matter. The diagram where the photon attaches to the lepton line also
  contributes.}
  \label{fig:dddiag}
\end{figure}

We now turn our attention to direct detection constraints. Skew-flavored 
DM can scatter off nuclei through loop diagrams involving the SM 
leptons, as shown in Fig.~\ref{fig:dddiag}. In particular, for any 
flavor of $\chi$ there will be two sets of diagrams that contribute to 
direct detection. These must be added at the amplitude level and then 
squared to obtain the total cross section. The calculation of each set 
proceeds in a manner that is identical to the case of conventional 
flavored DM, Eq.~(\ref{convFDM}). Therefore, in this paper we limit 
ourselves to writing down the final result and we refer the interested 
reader to \cite{Agrawal:2011ze} for additional details of the 
calculation.

In the case when $\chi_{\tau}$ is the lightest flavor, the
WIMP--nucleon cross section for direct 
detection is given by
\begin{align}
  \sigma_n
  =
  \frac{\mu_n^2 Z^2}{ A^2 \pi}
  \left(
  \frac{\lambda^2 e^2}{32\pi^2 m_\phi^2}
  X[e,\mu]
  \right)^2,
\end{align}
where $\mu_n$ is the reduced mass of the DM--nucleon system, and 
$Z$ and
$A$ are the atomic and mass number of the nucleus. Since the dark
matter dominantly scatters off protons, the WIMP--nucleon cross section scales
differently for different target nuclei.
$X[i,j]$ is defined for two flavors $i$ and $j$ as
 \begin{align}
  X[i,j]
  &\equiv  
  \left[
  1
  +\frac13 \log\frac{\Lambda_i^2}{m_{\phi}^2}
  +\frac13 \log\frac{\Lambda_j^2}{m_{\phi}^2}
  \right],
 \end{align}
 where $\Lambda_{\ell}$ represents the infrared cutoff in the loop 
calculation for the effective DM-photon coupling. This cutoff is well 
approximated by $m_{\ell}$, the mass of the corresponding lepton, unless 
$m_{\ell}$ is smaller than the momentum exchange in the process, which 
is of order 10~MeV. We therefore use $\Lambda_{\tau}=m_{\tau}$ and 
$\Lambda_{\mu}=m_{\mu}$, but $\Lambda_{e}=10~$MeV. 

\begin{figure*}[tp]
\includegraphics[width=0.45\textwidth]{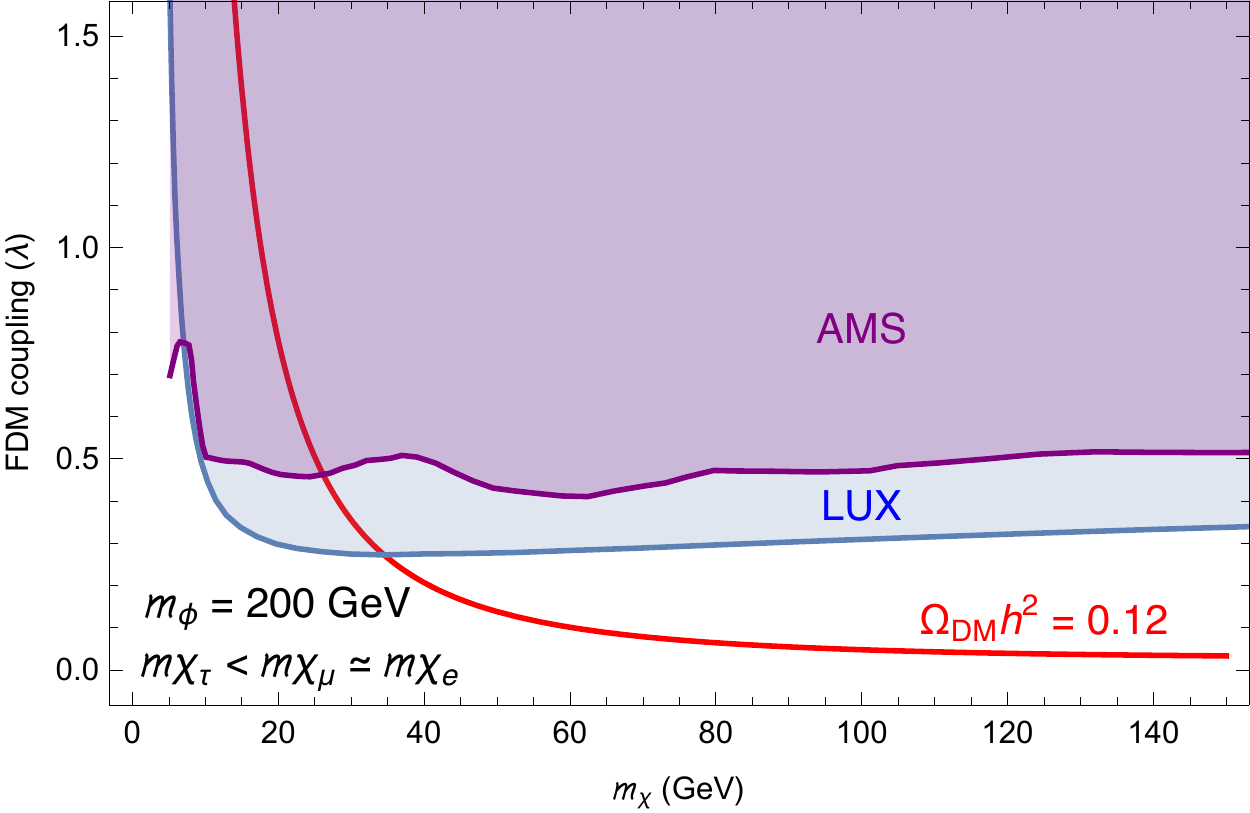}
\qquad
\includegraphics[width=0.45\textwidth]{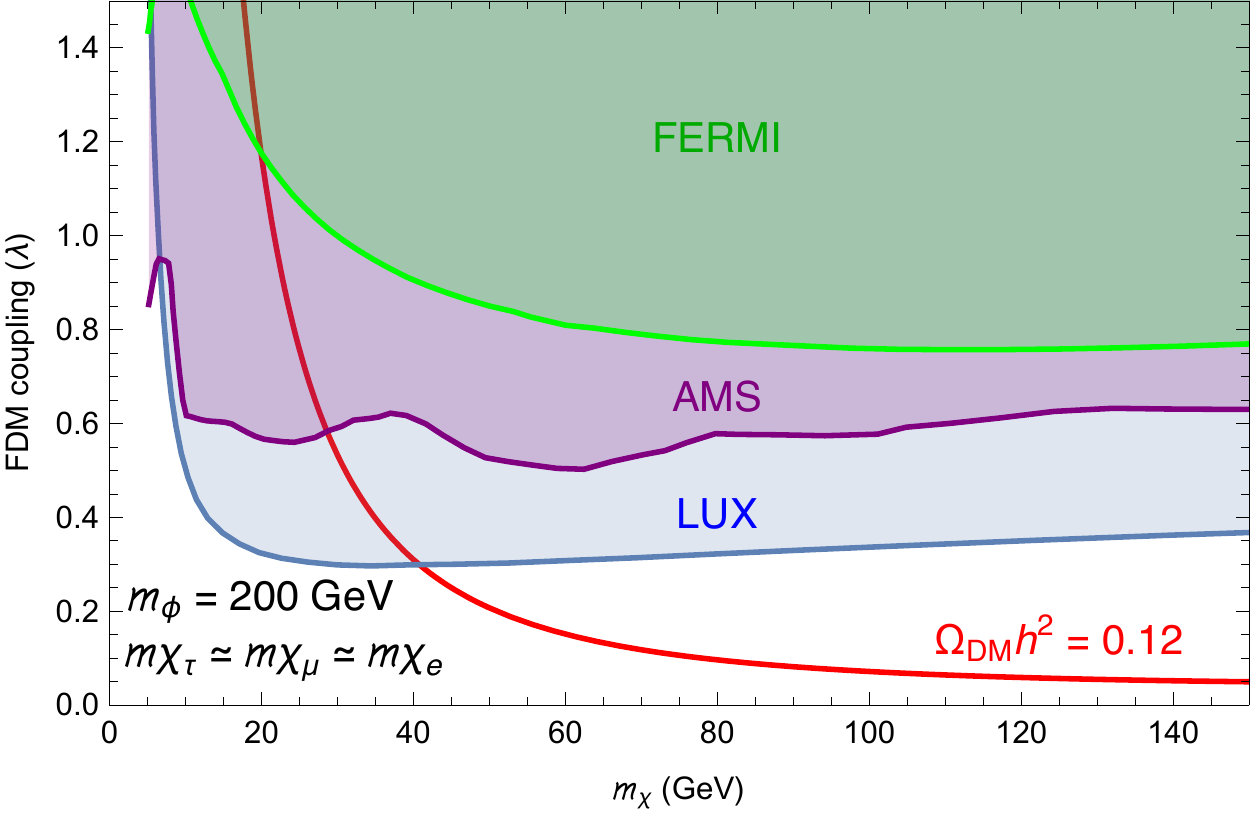}
\\
\vspace{20mm}
\includegraphics[width=0.45\textwidth]{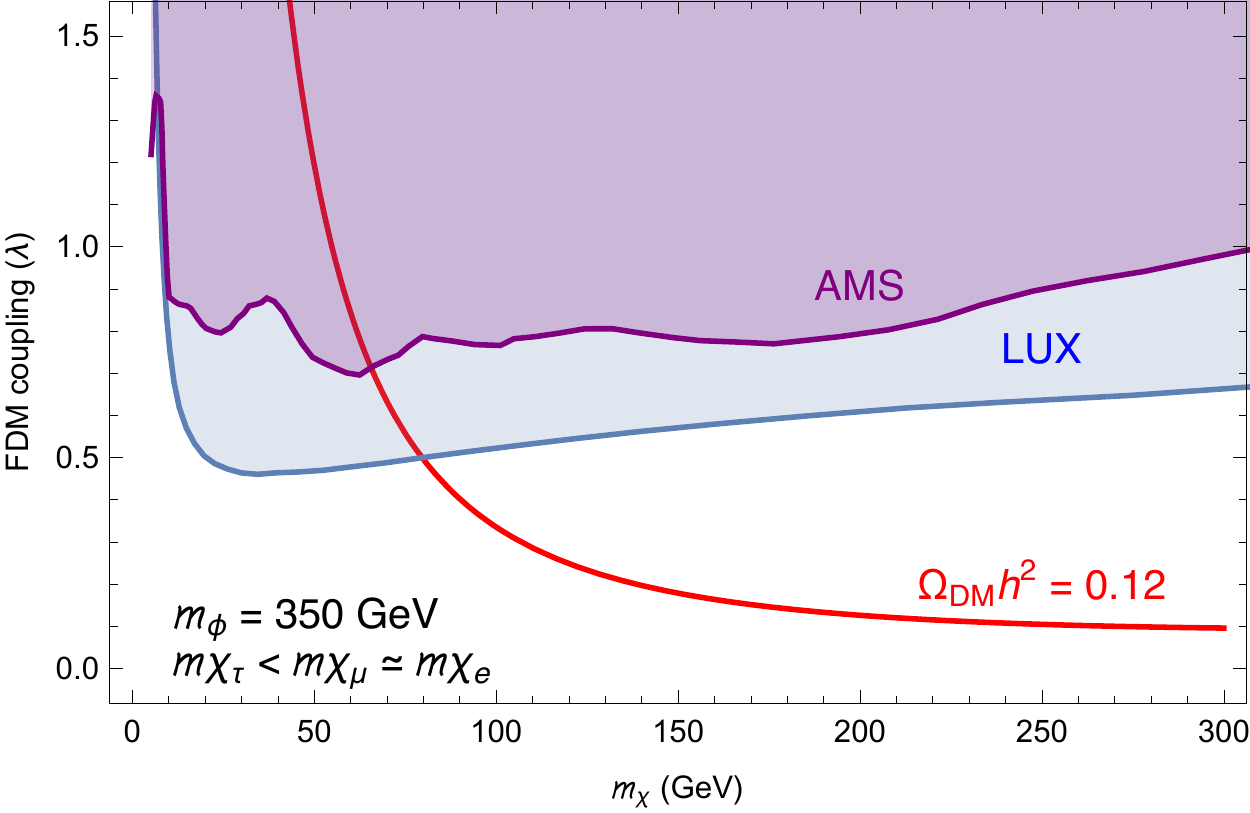}
\qquad
\includegraphics[width=0.45\textwidth]{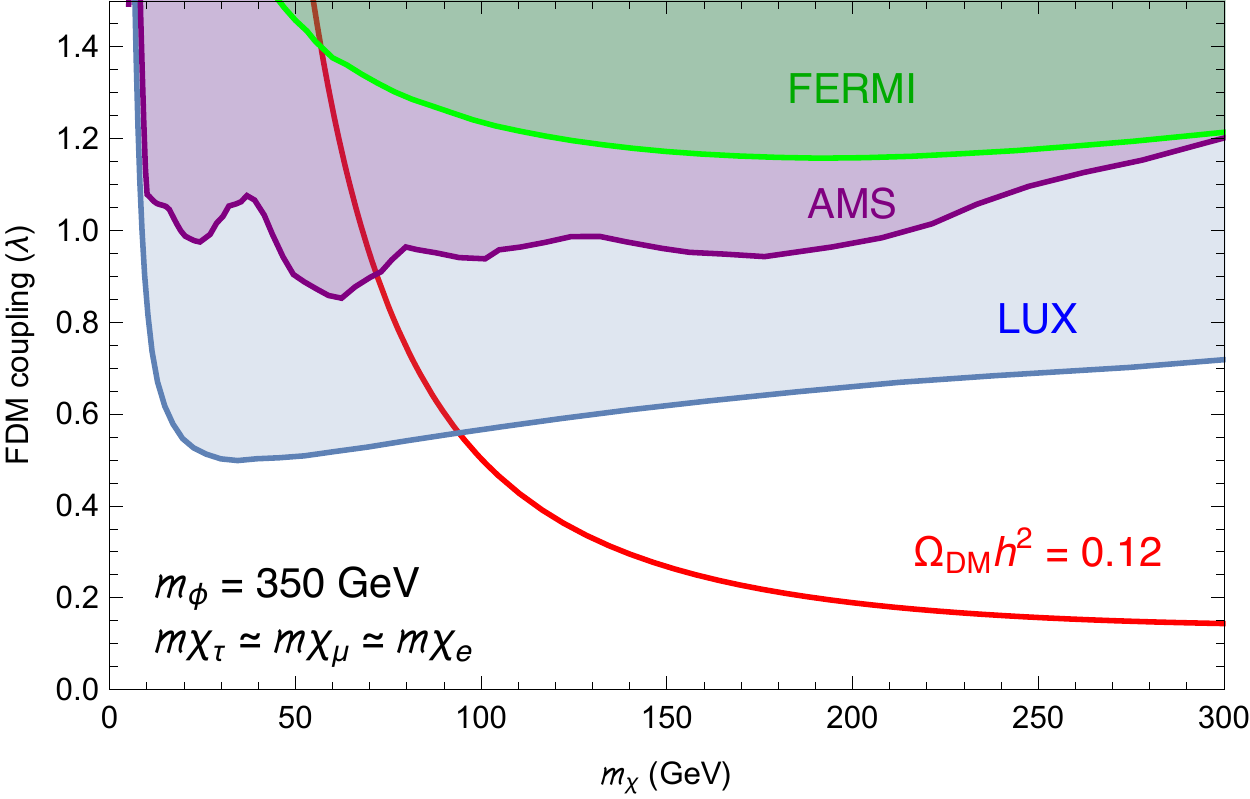}
\\
\vspace{20mm}
\includegraphics[width=0.45\textwidth]{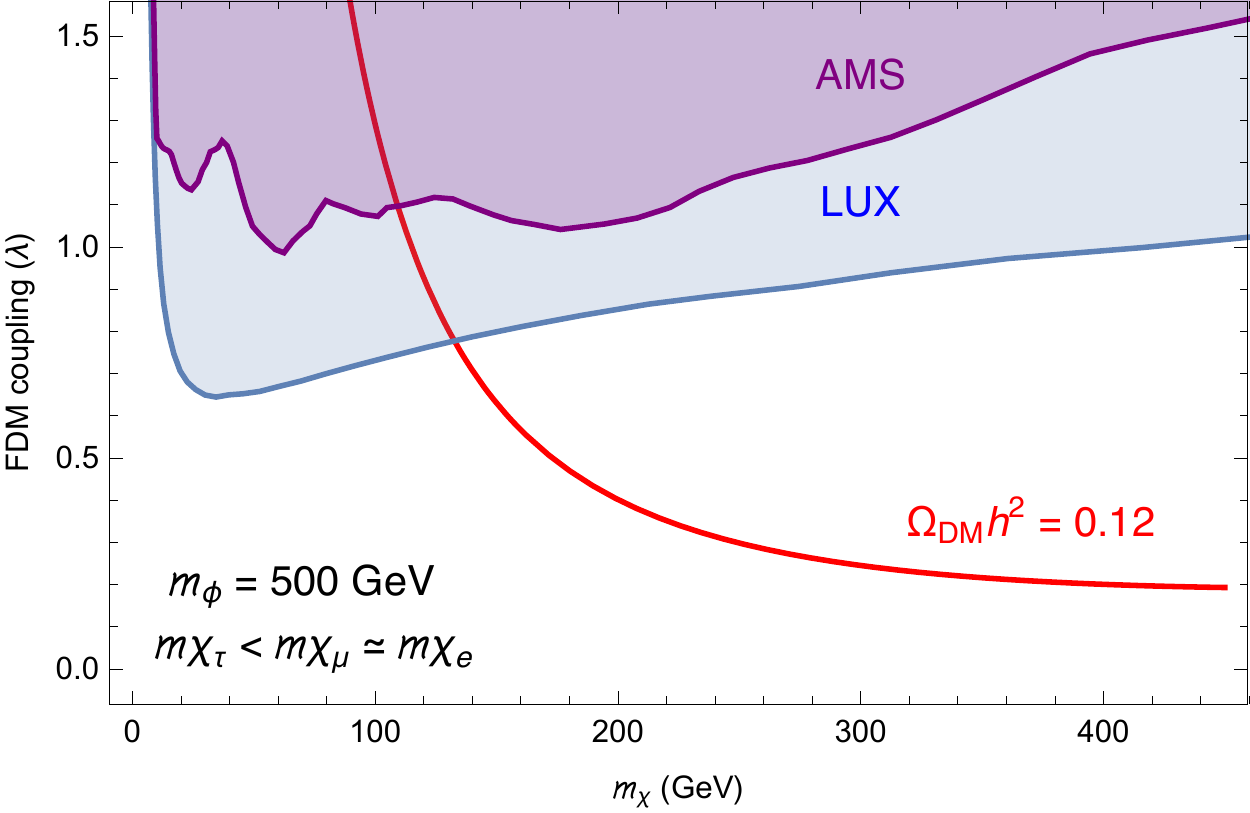}
\qquad
\includegraphics[width=0.45\textwidth]{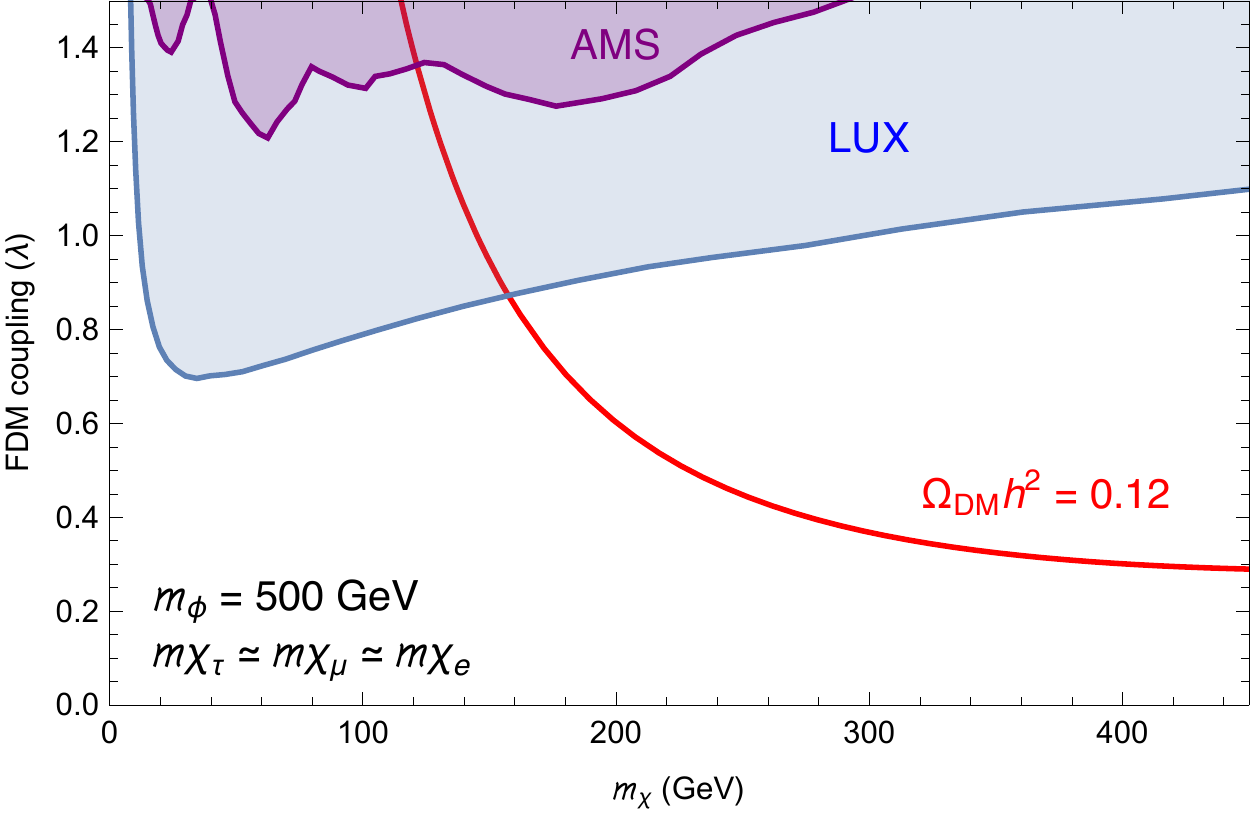}
\caption{For the cases of $\chi_{\tau}$ being the lightest flavor
(left plots), and for all three flavors degenerate (right plots), and
taking $m_{\phi}=\{200,350,500\}~$GeV (top to bottom), we illustrate
the region in
$m_{\chi}$ vs.  $\lambda$ parameter space where the correct relic
abundance is
obtained (red curve), the direct detection exclusion region from
LUX~\cite{Akerib:2015rjg}
(blue shaded region), the indirect detection exclusion region due
to AMS~\cite{Ibarra:2013zia}  (purple shaded region) and the exclusion
region due to FERMI gamma ray constraints~\cite{Tavakoli:2013zva}
(green shaded region).}
\label{fig:exclusion}
\end{figure*}

For the case where all three flavors of $\chi$ are degenerate and the
DM relic abundance today is made up of equal numbers of them, the
direct detection cross section is given by the average of their
individual scattering cross sections,
 \begin{eqnarray}
  \sigma_n
  &=&
  \frac{\mu_n^2 Z^2}{ A^2 \pi}
  \left(
  \frac{\lambda^2 e^2}{32\pi^2 m_\phi^2}
  \right)^2\times\nonumber\\
  &&\left(\frac{X[e,\mu]^{2}+X[e,\tau]^{2}+X[\mu,\tau]^{2}}{3}\right).
 \end{eqnarray}
 For $m_{\phi}=\{200,350,500\}~$GeV, we illustrate in 
Fig.~\ref{fig:exclusion} the region excluded by the LUX 
constraints~\cite{Akerib:2015rjg} for the two mass hierarchies of 
interest.

\subsection{Indirect Detection}

We now turn our attention to the limits from indirect detection, in 
particular from the AMS experiment for positrons~\cite{Ibarra:2013zia} 
and the FERMI experiment~\cite{Tavakoli:2013zva} for gamma rays. As in 
ref.~\cite{Agrawal:2015tfa}, we will make the simplifying assumption 
that the $\gamma$-ray flux is dominated by the decays of neutral pions 
$\pi^{0}\rightarrow\gamma\gamma$ produced after annihilation to 
$\tau^\pm$ pairs, and that photons that arise from the bremsstrahlung of 
electrons and muons can be neglected. We further assume that for a given 
$\chi$ mass, the positron bounds are dominantly sensitive to direct 
annihilation into positrons rather than to the secondary positrons that 
are produced in the decays of muons or taus, which exhibit a softer 
energy spectrum. While these approximations can certainly be improved 
upon, our results in Fig.~\ref{fig:exclusion} indicate that direct 
detection bounds always dominate over indirect detection bounds in this 
scenario, and therefore a full calculation of the positron and 
$\gamma$-ray fluxes would not affect the region of parameter space that 
is allowed.

Since DM annihilation proceeds dominantly through $s$-wave, the
annihilation cross sections relevant for indirect detection are the
same as those for relic abundance.
When $\chi_{\tau}$ is the lightest flavor, the annihilation cross
section is $\langle \sigma v \rangle_{relic}=2\langle\sigma
v\rangle_{0}$ from
Eq.~(\ref{eq:annxsecn}), and $e$ and $\mu$ pairs are produced with
equal probabilities in each annihilation event. With the assumptions
outlined above, we can then write down effective annihilation cross
sections that can be compared to the limits set by the FERMI and AMS
experiments
\begin{eqnarray}
  \langle\sigma v\rangle_{e^{+}\, eff}
  =\frac{1}{2}
  \langle\sigma v\rangle_{relic},\nonumber\\
  \langle\sigma v\rangle_{\gamma, eff}\approx 0 \; .
  \label{eq:IDtau}
\end{eqnarray}
Here $\langle\sigma v\rangle_{relic}$ is as defined in
Eq.~(\ref{eq:annxsecn}).

When all three flavors of $\chi$ are degenerate, and contribute equally 
to the total energy density, a similar analysis yields
\begin{eqnarray}
  \langle\sigma v\rangle_{e^{+}, eff}
  =
  \frac{1}{3}\langle\sigma v\rangle_{relic}\nonumber
  \\
  \langle\sigma v\rangle_{\gamma, eff}
  =
  \frac{1}{3}\langle\sigma v\rangle_{relic} \,  .
  \label{eq:ID3}
\end{eqnarray}
Here $\langle\sigma v\rangle_{relic} = \frac43 \langle \sigma v \rangle_0$
is the annihilation rate defined in Eq.~(\ref{eq:effxsec3}).
As noted earlier, the annihilation cross section for two DM particles 
of the same flavor yields two possible final 
states with equal probabilities. For a different flavor channel we 
obtain a unique final state. For example, a $\bar{\chi}_{e}\chi_{e}$ 
initial state gives rise to equal amounts of $\mu\mu$ and $\tau\tau$ 
final states, while annihilations of $\chi_{e}$ with $\chi_{\mu}$ lead 
to $e\mu$ final states only. All these considerations have been taken 
into account in deriving Eq.~(\ref{eq:ID3}).

For $m_{\phi}=\{200,350,500\}~$GeV, we illustrate in 
Fig.~\ref{fig:exclusion} the region excluded by positron and 
$\gamma$-ray constraints in the two mass hierarchies of interest. It 
should be pointed out that the limits quoted by AMS and FERMI are 
calculated for a Majorana fermion, therefore in plotting 
Fig.~\ref{fig:exclusion} we include the necessary factor of 2 to 
compare with predictions for $\chi$, which is a Dirac fermion.

\subsection{Other constraints}

There are other potential constraints that arise from the interaction
in Eq.~(\ref{epsint}). For example, this interaction can lead to
corrections to the anomalous magnetic
moment of the muon~\cite{Bai:2014osa,Chang:2014tea,Agrawal:2014ufa}.
However, the $(g-2)$ bound constrains the mediator masses $m_\phi \sim
100$ GeV for coupling $\lambda\sim1$. We see from
Fig.~\ref{fig:exclusion} that direct detection
constraints are more stringent than those from $(g-2)$ of the muon.
There are also potential constraints from monophoton production at
LEP~\cite{Fox:2011fx}. Again, as shown in~\cite{Agrawal:2014ufa},
these bounds are weaker than the limits from direct detection.

For very light masses $m_\chi\lesssim 5$ GeV, the direct and indirect
detection constraints are substantially weakened, and the $(g-2)$ and
monophoton constraints might play a role in this region. However, as
we see from
Fig.\ref{fig:exclusion}, this region is not compatible with dark
matter relic abundance, and hence would require a more complicated
cosmological history to be viable.

\subsection{Collider Signals}

We now consider the collider signatures associated with this class of 
models at the LHC. Since $\chi_{e}$ and $\chi_{\mu}$ are expected to be 
highly degenerate, decays from one to the other may not be kinematically 
allowed. Even in a scenario where these decays are allowed, the 
resulting charged leptons are extremely soft, and would be challenging 
to detect in an LHC environment. For the purposes of the following 
discussion, we will therefore assume that these leptons are not 
detected.

\begin{figure}[tp]
  \begin{center}
    \includegraphics[width=0.3\textwidth]{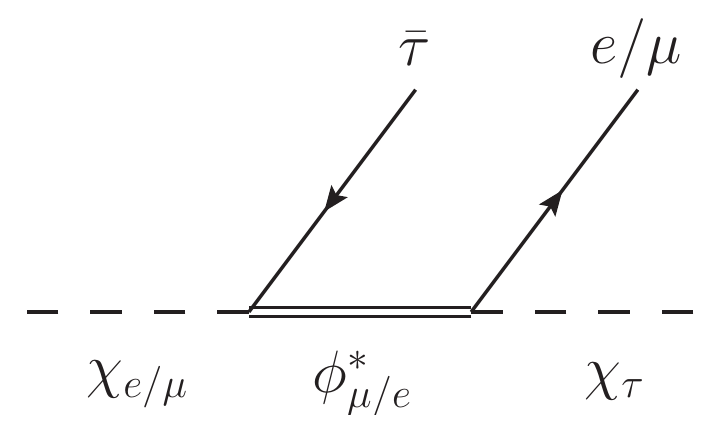}
  \end{center}
  \caption{Feynman diagram for the decay of $\chi_{e,\mu}$ to $\chi_{\tau}$.}
  \label{fig:chidecay}
\end{figure}

On the other hand, in the case when $\chi_{\tau}$ is sufficiently split 
from the other two flavors, the leptons produced in the decays can indeed be 
detected. As shown in figure~\ref{fig:chidecay}, if $\chi_\tau$ is the lightest flavor, $\chi_e$ and 
$\chi_\mu$ will decay down to it, these decays being accompanied by a 
tau and an electron in the case of of $\chi_e$, and by a tau and a muon 
in the case of $\chi_{\mu}$.

\begin{figure}[tp]
  \begin{center}
    \includegraphics[width=0.3\textwidth]{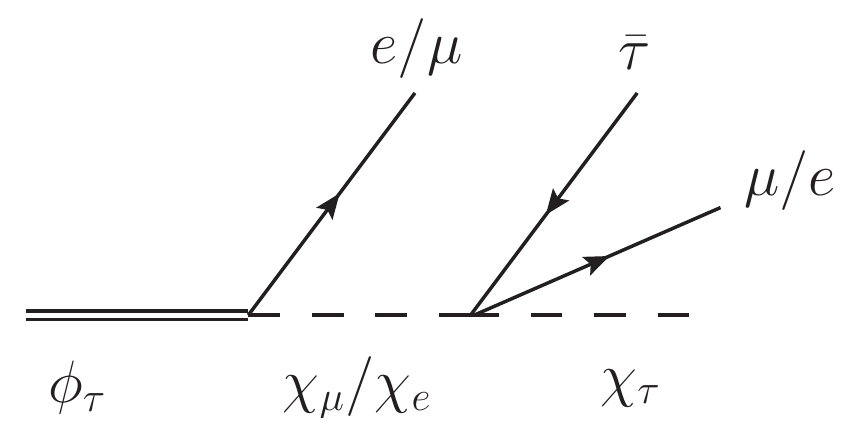}
  \end{center}
  \caption{Feynman diagram for the decay of $\phi_{\tau}$.}
  \label{fig:phitaudecay}
\end{figure}

With this in mind, let us consider the scenario in which $\chi_{\tau}$
is the lightest flavor. The mediators $\phi_e$, $\phi_{\mu}$ and
$\phi_\tau$ can be pair-produced in colliders through an off-shell
photon or $Z$. As shown in figure~\ref{fig:phitaudecay}, each $\phi_{\tau}$ decays either to $\chi_e + \mu$ or
to $\chi_\mu + e$, followed by the decay of $\chi_e$ or $\chi_\mu$ to
$\chi_\tau$. We see that each $\phi_\tau$ eventually decays down to
$\chi_{\tau}$ accompanied by an electron, a muon and a tau. It follows
that events where $\phi_\tau$ is pair-produced are characterized by two
electrons, two muons, two taus and missing energy.

\begin{figure}[tp]
  \begin{center}
    \includegraphics[width=0.18\textwidth]{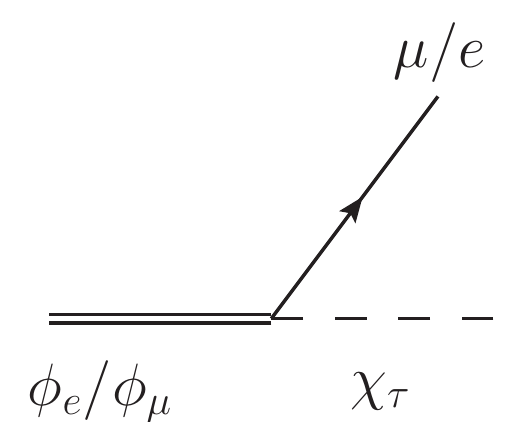}\\
    \includegraphics[width=0.3\textwidth]{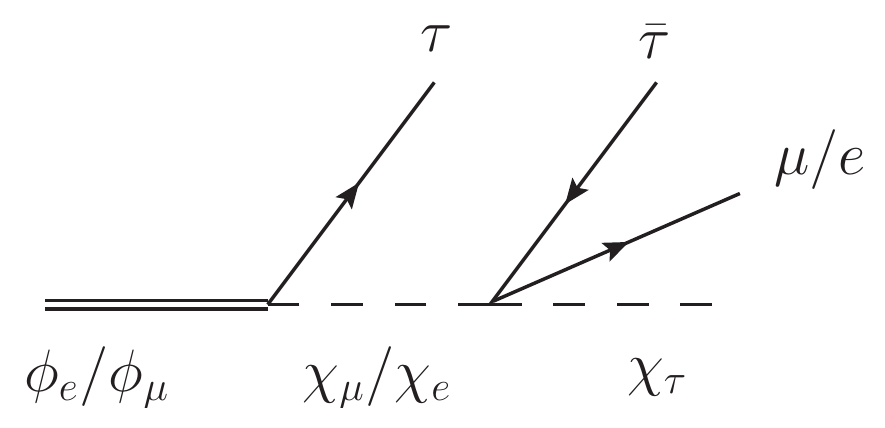}
  \end{center}
  \caption{Feynman diagrams for the short and long decay modes of $\phi_{e,\mu}$.}
  \label{fig:phiedecay}
\end{figure}

Each $\phi_e$, on the other hand, can decay directly to $\chi_\tau$
accompanied by a single muon. Alternatively, it can decay down to
$\chi_\mu + \tau$, followed by the decay of $\chi_\mu$ to $\chi_\tau$,
which is accompanied by a muon and a tau. We label the direct decay to
$\chi_\tau$ as the short chain, and the cascade decay as the long
chain. These are shown in figure~\ref{fig:phiedecay}.  It follows that when $\phi_e$ is pair-produced, we can obtain
an event with two short chains, with a short chain and a long
chain, or with two long chains, each containing two muons and missing energy, and zero, two or four taus, respectively. Similarly,
when $\phi_\mu$ is pair-produced, we can obtain an event with two
electrons and missing energy, and zero, two or four taus.

Since all these final states are produced with comparable branching
fractions, the more spectacular ones will dominate in setting the
discovery potential, since the corresponding backgrounds will be
negligible for all intents and purposes. In particular, a search
strategy can be set up to look for the 2$e$-2$\mu$ + MET final state,
with the additional requirement that at least one good $\tau$
candidate be present in the event. Given the smallness of SM
backgrounds for such a final state, as long as the signal results in
${\mathcal O}(10)$ events after detector and analysis efficiencies are
taken into account, discovery-level significance should be easy to
obtain. Furthermore, once an excess of this kind is observed, the
additional 2$\ell$-4$\tau$ final state with comparable numbers of
signal events can be used as a cross-check to not only bolster the
discovery significance, but as a smoking gun to distinguish
skew-flavored DM from any other potential explanation of the excess.

The best existing constraints on the $4\ell$ final states come from the multilepton searches by ATLAS and CMS~\cite{Chatrchyan:2014aea,Aad:2014hja}. Note that the $4\ell$ states are populated by $\phi_{\tau}$ pair production, and always result in two opposite-sign same flavor pairs (which may have the same or different flavors), plus $\tau$'s. Thus they populate the OSSF2 category in CMS' notation. Since there is no significant anomalous excess in these categories, the mass of $\phi_{\tau}$ should be heavy enough such that the expected number of events with 20~fb$^{-1}$ of luminosity at 8~TeV is less than one. This pushes the $\phi_{\tau}$ mass to roughly 500~GeV. As can be seen in figure~\ref{fig:exclusion} (lower left plot), there is a region of parameter space where all constraints can be satisfied. Note that at 13~TeV the luminosity function for a $q$-$\bar{q}$ initial state at these masses is larger than that at 8~TeV by a factor of 5 or so, and since the backgrounds are so low, even a small number of events observed in run II may quickly become significant.

When all three flavors are degenerate and long-lived on cosmological 
scales, the collider phenomenology is much simpler. In particular, any 
flavor of $\phi$, once produced, will have a 50\% chance to decay to one 
of the other two flavors of $\chi$ along with a SM lepton. Any $\chi$ so 
produced will simply leave the detector as missing energy without 
further decaying. Therefore, once the contributions from the three 
$\phi$ flavors are added up, this means that any event will result in a 
pair of opposite sign, flavor uncorrelated leptons, and additional 
missing energy. While these final states are not as background-free as 
the ones discussed above, they are similar to the final states that 
occur in other extensions of the Standard Model, such as chargino pair 
production followed by $\chi^{\pm}\rightarrow\ell^{\pm}\nu\chi^{0}$ in 
supersymmetric models. As in those theories, the existence of additional 
missing energy in the final state makes it possible for kinematic cuts 
such as $m_{T2}>m_{W}$ to be used to enhance the signal to background 
ratio. We therefore expect that discovery is still possible as long as 
the $\phi$ masses are low enough to result in observable production 
cross sections.

Note that even in the degenerate case where multilepton final states do not arise, there is still something unique about the flavor correlations that could in principle be used as a smoking gun signature to identify the underlying skew-flavored DM model. In particular, since any given $\phi$ flavor can decay to two, but not all three possible lepton (and $\chi$) flavors, and since pair production always starts with a same-flavor $\phi$-pair, the ratio of the number of same-flavor over different-flavor dilepton events in the final state is 1 (of course, within the same-flavor and the different-flavor final states, all flavor combinations are equally likely as long as the $\phi$ flavors are degenerate). This is in contrast to both SM backgrounds such as $W^{+}W^{-}$ and $t$-$\bar{t}$, as well as to flavor-uncorrelated beyond-the-SM signals such as chargino pair production in SUSY with decays to the LSP. In all those cases, since all three lepton flavors can be produced on either side of the event, the same-flavor to different-flavor ratio in the final state lepton pairs is 1/2. Therefore, if the signal can be purified by cutting on a kinematic variable such as $m_{T2}$, then this ratio can be used to check the skew-flavored DM hypothesis.

For the degenerate case, the best existing constraints come from ATLAS and CMS searches for supersymmetry with two opposite sign leptons~\cite{Aad:2014vma,Chatrchyan:2013kha}. Since this is a less exotic state than $4\ell+\tau$'s, the backgrounds are larger, and even though there is no statistically significant excess, the data can accommodate ${\mathcal O}(10)$ events, see for instance table 5 of the ATLAS reference. Considering that all three flavors of $\phi$-pairs can contribute to this final state, this results in mass bounds on the $\phi$ particles of roughly 350~GeV. As can be seen in the middle right plot of figure~\ref{fig:exclusion}, there is again a region of parameter space that is consistent with all constraints. Similar to the degenerate case, the cross section at 13~TeV is larger than that at 8~TeV by roughly a factor of 5, so if the $\phi$ masses are not far above the bound, signal events should be observable with a moderate luminosity in run II. Of course, unlike the $4\ell$ final state where due to the low backgrounds kinematic cuts can be very low in searches, for the $2\ell$+MET channel one has to rely on cuts on variables such as $m_{T2}$, and therefore to get precise bounds or to map out discovery regions one has to evaluate signal efficiencies using Monte Carlo simulation and reconstruct the statistical procedures used in the ATLAS and CMS analyses. This is beyond the scope of this paper which aims to describe the general characteristics of the skew-FDM scenario, but it will be taken up in future work.

\section{Alternative Realizations of Skew-Flavored DM}

Although our primary focus has been on the case in which DM couples to 
the right-handed leptons of the SM, it is straightforward to construct 
theories of skew-flavored DM in which the $\chi$ fields couple to the 
left-handed leptons, or to the quarks. Consider first the case in which 
the DM particles couple to the left-handed leptons. The interaction term 
then takes the form
 \begin{align}
\lambda_{ABC} \chi^A L^B \phi^C \; \; + {\rm h.c.} \; ,
 \end{align}
 where $A$, $B$ and $C$ represent SU(3$)_{L}$ flavor indices. The $\chi$ 
fields are taken to be SM singlets, while the mediator fields $\phi$ 
transform as doublets under the SM SU(2$)_{\rm L}$ gauge symmetry.
   
This scenario gives rise to signals that are qualitatively quite similar 
to those we have studied. However, since the mediators are now charged 
under the weak interactions, their production cross section at the LHC 
is significantly larger. In addition, some of the final state particles 
produced in the decay chains will now be neutrinos rather than charged 
leptons. Therefore, although the characteristic signatures still involve 
leptons and missing energy, a typical signal event is expected to have 
larger missing energy, but fewer charged leptons in the final state, 
than in the scenario we have focused on.

Theories in which the DM fields couple to the SM quarks tend to be more 
severely constrained, both from direct detection experiments and from existing searches at the 
LHC. The interaction term takes the form
 \begin{align}
\lambda^{ijk} \chi_i D^c_j \phi_k \; \; + {\rm h.c.}
 \end{align}
 for the case of coupling to the right-handed down-type quarks, and
 \begin{align}
\lambda^{ijk} \chi_i U^c_j \phi_k \; \; + {\rm h.c.}
 \end{align}
 for the case of coupling to the right-handed up-type quarks. Here $i$, 
$j$ and $k$ represent SU(3$)_{D}$ or SU(3$)_{U}$ flavor indices 
respectively. In the case of coupling to left-handed quarks, the 
interaction term becomes
 \begin{align}
\lambda_{ABC} \chi^A Q^B \phi^C \; \; + {\rm h.c.} \; ,
 \end{align}
 where $A$, $B$ and $C$ represent SU(3$)_{Q}$ flavor indices.  The 
$\chi$ fields are once again taken to be SM singlets, while the mediator 
fields $\phi$ are now charged under the SM color group.

If the lightest DM particle carries the flavor quantum numbers of any of 
the down, strange or bottom quarks, the skewed flavor structure admits 
scattering at tree level off nuclei, leading to stringent bounds from 
direct detection. Similar considerations apply to the top- and 
charm-flavored scenarios. The lone exception is the scenario in which 
the lightest DM particle carries up-flavor. Since the nucleon has no 
significant charm or top content, in this case the dominant direct 
detection signal is loop suppressed.

\begin{figure}[tp]
  \begin{center}
    \includegraphics[width=0.3\textwidth]{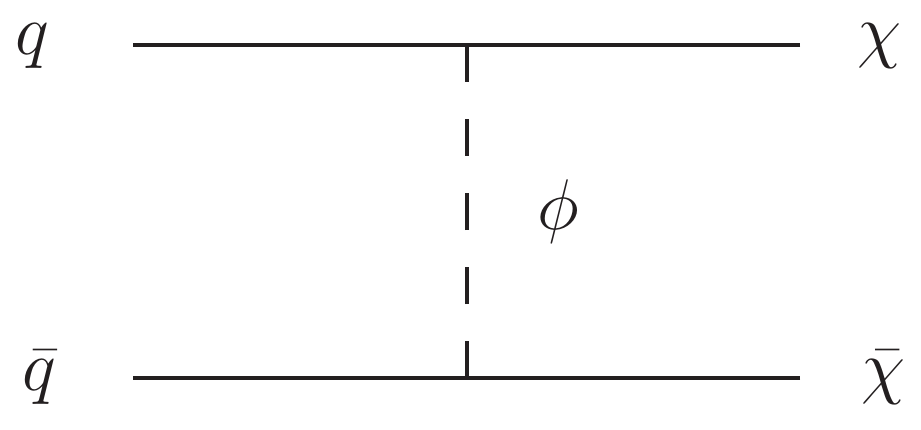}
  \end{center}
  \caption{Feynman diagram for pair production of DM particles through $\phi$-exchange. Flavor indices have been suppressed.}
  \label{fig:chichiprod}
\end{figure}

\begin{figure}[tp]
  \begin{center}
    \includegraphics[width=0.3\textwidth]{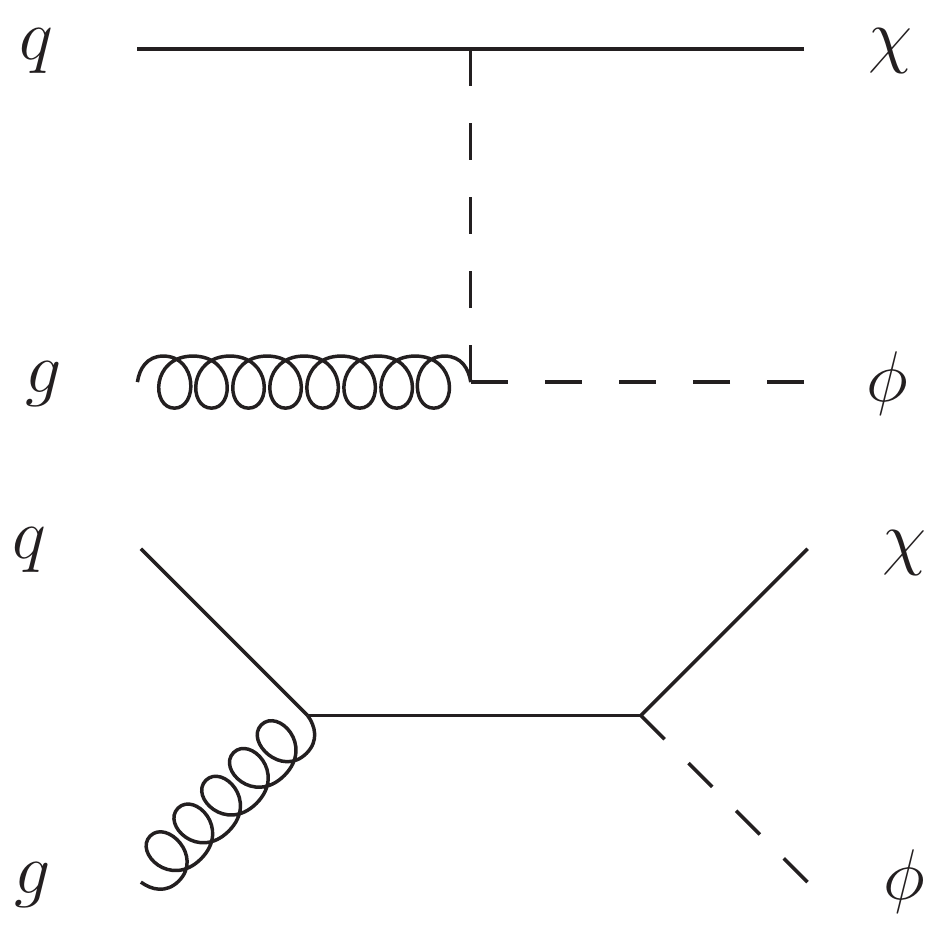}
  \end{center}
  \caption{Feynman diagrams for associated $\chi$-$\phi$ production. Flavor indices have been suppressed.}
  \label{fig:phichiprod}
\end{figure}

The collider phenomenology of the quark-flavored scenario is also very 
different from the lepton-flavored case. In particular, qualitatively 
distinct types of production mechanisms can now give rise to signal 
events. Since the mediators now carry SM color, their pair production 
cross section at the LHC is much larger than in the case we have 
studied, and the corresponding limits from searches for jet events with 
missing energy are expected to be very strong. In addition, there are 
now events involving direct pair production of DM particles through 
$\phi$ exchange, as shown in Fig.~\ref{fig:chichiprod}. Provided the 
mass splittings are large enough, decays of the heavier DM flavors will 
result in observable signatures involving jets and missing energy. In 
regions of parameter space in which the mediators are much heavier than 
the DM fields, these events are expected to give rise to the dominant 
signal. It is now also possible to directly produce a DM particle in 
association with a mediator, as shown in Fig.~\ref{fig:phichiprod}. This 
class of events also gives rise to jets plus missing energy signatures. 
Overall, the bounds on quark-flavored scenarios are expected to be much 
stronger than in the lepton-flavored case we have focused on.

\section{Conclusions}

Flavored DM is a simple possibility consistent with the WIMP miracle and 
the non-observation of flavor violation beyond the SM. Null results in 
current direct detection experiments further motivate flavor-specific 
couplings of the DM with SM matter. In this paper we presented a novel 
skewed flavor structure in flavored DM models that is consistent with 
MFV. We studied the phenomenology of a benchmark model in which the DM 
has contact interactions with right-handed leptons, and the interactions 
are fully antisymmetric in flavor space. Unlike the conventional 
flavored DM scenario, the skew-flavored setup includes three flavors of 
the mediator. Depending on the mass splittings between the DM flavors, 
the cosmological relic abundance today may be due to a single flavor, or 
a mixture of two or even three nearly-degenerate flavors.

We have found that in large regions of parameter space the DM can arise 
as a thermal relic while remaining consistent with the null results of 
direct and indirect detection experiments. This allowed parameter space 
also includes regions where the mediator is light enough to be 
pair-produced at the LHC with a potentially observable cross section. 
When that happens, the mediators decay to the lightest DM flavor, 
frequently resulting in cascade decays, and the associated collider 
signatures involve multi-lepton final states with additional missing 
energy. An order one fraction of the signal events can therefore produce 
final states with essentially vanishing SM backgrounds, making discovery 
a possibility even for low production cross sections. Furthermore, the 
flavor pattern in the signal events is quite distinctive, and can 
potentially be used to distinguish this class of theories from other 
flavored and unflavored DM models. The detailed collider phenomenology 
of this scenario will be the subject of a future study.

\acknowledgments 
 Fermilab is operated by Fermi Research Alliance, LLC under Contract No. 
De-AC02-07CH11359 with the United States Department of Energy. ZC is 
supported by the NSF under grant PHY-1315155. ECFSF thanks the 
University of Maryland and NASA Goddard Space Flight Center for the 
hospitality while this work was being completed and FAPESP for full 
support under contracts numbers 14/05505-6 and 11/21945-8. The research 
of CK is supported by the National Science Foundation under Grants No. 
PHY-1315983 and No. PHY-1316033. The work of PA was supported in part by 
NSF grants PHY-0855591 and PHY-1216270. PA would also like to thank the 
Aspen Center for Physics, which is supported by National Science 
Foundation grant PHY-1066293.

\bibliography{fdmskew.bib}

\end{document}